# IMPACTO DEL DESORDEN EN LOS ESTADOS CUÁNTICOS DE DOS FOTONES GENERADOS EN ARREGLOS DE GUÍAS DE ONDA NO LINEALES

Jefferson Delgado-Quesada y Edgar A. Rojas-González[*]

Escuela de Física, Universidad de Costa Rica, San José, Costa Rica
Centro de Investigación en Ciencia e Ingeniería de Materiales, Universidad de Costa Rica, San José, Costa Rica

**Abstract**

In an array of nonlinear waveguides, quantum states can be generated from classical states through spontaneous parametric down-conversion of photons. This work simulates and analyzes the effect of disorder on light propagation and its quantum correlations, implementing disorder in three system parameters: coupling, injection amplitude, and injection phase. It is determined that when light is injected into only one waveguide, disorder in the coupling increases localization and reduces dispersion. Moreover, in this case, propagation tends to remain ballistic, a characteristic feature of waveguide arrays. Conversely, disorder in the injection amplitude and phase tends to delocalize the wave function, with the latter having a more pronounced effect. Finally, it is observed that quantum correlations, obtained from the correlation matrix, are robust in the presence of disorder, particularly the null elements.

**Resumen**

En un arreglo de guías de onda no lineales se pueden generar estados cuánticos a partir de estados clásicos mediante la conversión paramétrica descendente espontánea de fotones. En este trabajo se simula y analiza el efecto del desorden en la propagación de la luz y sus correlaciones cuánticas, implementando desorden en tres perfiles del sistema: acoplamiento, amplitud de inyección y fase de inyección. Se determina que, cuando se inyecta solamente una guía de onda, el desorden en el acoplamiento aumenta la localización y disminuye la dispersión. Además, en este caso se preserva la propagación con tendencia a ser balística, lo cual es característico de arreglos de guías de onda. Contrariamente, el desorden en la amplitud y fase de inyección tienden a deslocalizar la función de onda, siendo el efecto mayor en el último caso. Finalmente, se observa que las correlaciones cuánticas, obtenidas a partir de la matriz de correlación, son robustas ante la presencia de desorden, en particular los elementos nulos.

**Palabras clave:** Desorden, fotónica integrada, óptica cuántica, QuTiP.

# I. INTRODUCCIÓN

La fotónica integrada tiene la capacidad de generar estados entrelazados, manipular estados cuánticos y mantener un bajo nivel de ruido. Por lo tanto, esta es una plataforma robusta para el desarrollo de tecnologías cuánticas. Una de sus características más importantes es que permite mantener la coherencia de una forma más sencilla en comparación con sistemas a base de materia (Wang et al., 2020)—es decir, no requiere temperaturas extremadamente bajas o alto vacío como en el caso de los últimos. De hecho, la fotónica integrada ha sido utilizada en varias aplicaciones en el contexto de tecnologías cuánticas—

EDGAR.ROJASGONZALEZ@ucr.ac.cr

por ejemplo, en comunicación, computación, procesamiento de información y simulación de sistemas físicos o químicos (Laurent Labonté et al., 2024; Luo et al., 2023; Wang et al., 2020).

Un dispositivo sencillo para estudiar el alcance de la fotónica integrada es un arreglo de guías de onda. En un sistema homogéneo—aquel en donde todas las guías de onda son idénticas y poseen el mismo acoplamiento—un par de fotones experimenta interferencia y se observan efectos no clásicos, como aglomeración o una probabilidad nula para una medida de coincidencia (Bromberg et al., 2009). Este sistema ha sido ampliamente utilizado para el estudio del *quantum random walk*, cuyo mayor reto es fácilmente superado en un arreglo de guías de onda (Perets et al., 2008)—es decir, mantener la coherencia en un sistema de gran escala.

En un contexto clásico, se le llama *random walk* a una partícula que se mueve aleatoriamente a posiciones adyacentes en un arreglo con una probabilidad dada por una distribución gaussiana y una distancia promedio con respecto a su punto inicial proporcional a la raíz cuadrada del tiempo. El *quantum random walk*, a diferencia del *random walk* clásico, se propaga con una rapidez directamente proporcional al tiempo, efecto denominado propagación balística (Perets et al., 2008). Los pasos durante un *random walk* pueden ocurrir con un tiempo discreto, es decir, los movimientos solo se producen cada cierto periodo, o con un tiempo continuo, para el cual la partícula puede realizar un paso en cualquier instante (Peruzzo et al., 2010).

Un arreglo de guías de onda exhibe específicamente un *quantum random walk* de tiempo continuo caracterizado por no evolucionar a un estado estacionario, a diferencia del caso clásico; esto lo vuelve altamente dependiente de las condiciones iniciales (Peruzzo et al., 2010). Este efecto es importante en ciencia de la computación dado que es una herramienta útil en algoritmos utilizados para resolver problemas de conteo o relacionados con grandes conjuntos de estructuras (Kempe, 2003).

Un arreglo de guías de onda lineal permite modificar los estados cuánticos de la luz, pero no es capaz de aumentar el grado de entrelazamiento entre los fotones (Solntsev et al., 2012). Un arreglo no lineal trasciende este límite ya que, en este, un estado inicial clásico sí puede evolucionar a un estado no clásico. En este arreglo de guías de onda, se dice que las guías están acopladas, ya que puede ocurrir efecto túnel entre ellas. Además, el grado de no linealidad del dispositivo está relacionado con la potencia de la inyección inicial y la susceptibilidad eléctrica del material de las guías de onda (Rai y Angelakis, 2012; Solntsev et al., 2012). Este sistema es entonces adecuado para el estudio de fenómenos no lineales y la implementación de desorden en sus parámetros; específicamente, su impacto en la propagación de la luz y las correlaciones cuánticas entre los modos individuales (las guías de onda) del arreglo.

La principal aplicación de los arreglos de guías de onda no lineales es la generación de estados cuánticos específicos a través del diseño del dispositivo. Por ejemplo, Barral et al. (2020a) exploraron, mediante el análisis de soluciones analíticas válidas en ciertos regímenes, la producción de fotones entrelazados y otros efectos cuánticos para diversas condiciones iniciales. Además, examinaron el efecto de modificar ciertos parámetros, como la fuerza de acoplamiento, la estrategia de medición o la fase y amplitud de la inyección inicial de fotones.

Introducir desorden en un sistema puede producir efectos como la interferencia por dispersión, resultando en localización espacial. Por ejemplo, una distribución de energía aleatoria en una red causa un fenómeno conocido como localización de Anderson, donde la función de onda se concentra en una región específica, lo que aumenta la probabilidad de encontrar la partícula ahí (Anderson, 1958). Karamlou et al. (2022) exploraron esto en una red de enlace fuerte constituida por *qubits*. En el caso de los electrones, por ejemplo, la localización de Anderson disminuye la conductividad, por lo que este efecto es interesante para el estudio de transiciones de fases conductoras a fases no conductoras en un sólido.

Las publicaciones relacionadas con el impacto del desorden en un arreglo de guías de onda no lineales no son abundantes. Existen estudios sobre sus aplicaciones en algoritmos de búsqueda cuántica (Hamilton et al., 2014) y protección topológica de estados cuánticos (Blanco-Redondo et al., 2018); sin embargo, dichos trabajos no profundizan en el impacto del desorden, sino que se enfocan en defectos locales.

Karamlou et al. (2022) exploraron el efecto del desorden en una red de enlace fuerte; sin embargo, tanto el tipo de desorden como el sistema son distintos a los que se estudiarán en este trabajo. En el contexto de arreglos de guías de onda Kokkinakis et al. (2024) han estudiado sistemas que exhiben una asimetría en el acoplamiento, es decir, un acoplamiento que depende de la dirección del salto de los fotones entre las guías de onda, aunque su trabajo no abarca arreglos no lineales. Por otro lado, Bai et al. (2016) sí estudian arreglos no lineales, pero solamente analizan valores puntuales de desorden y se limitan a inyecciones de una o dos guías de onda.

La ausencia de un análisis profundo del desorden en un arreglo no lineal, en conjunto con la relevancia de dicho dispositivo en áreas como información cuántica, computación y comunicación, justifica la importancia de este trabajo y la exploración del efecto del desorden en la generación de estados cuánticos particulares. Por lo tanto, en este trabajo se analiza el efecto del desorden en la propagación de la luz y las correlaciones cuánticas generadas en arreglos de guías de ondas no lineales. Para esto se realizan simulaciones numéricas de la propagación de la luz y se describe el efecto del desorden en la generación de estados cuánticos particulares.

**Arreglo de guías de onda no lineales**

El proceso no lineal que se va a estudiar en este trabajo en relación con arreglos de guías de onda es conocido como conversión paramétrica descendente espontánea. En este, un fotón inyectado en el arreglo se convierte en dos fotones hijos con menor frecuencia, según los requerimientos de conservación de energía y momento lineal. Se le llama espontáneo debido a la naturaleza aleatoria de los eventos y paramétrico porque la luz no afecta el estado cuántico del material (Zhang et al., 2021).

En este trabajo se estudia un arreglo de $N$ guías de onda no lineales idénticas con una susceptibilidad cuadrática $\chi^{(2)}$ con acoplamiento solamente entre las guías adyacentes, donde la luz se propaga a lo largo de la dirección $z$ hasta una distancia $L$, llamada longitud de la zona de interacción—región en donde la luz experimenta acoplamiento y efectos no lineales. Existen dos tipos de acoplamiento: el acoplamiento lineal entre las guías $j$ y $j+1$ caracterizado mediante $C_j$ y la fuerza de acoplamiento $C_0$, y el acoplamiento no lineal efectivo de la guía $j$ caracterizado mediante $\eta_j = g\alpha_j$, donde $g$ es la constante no lineal, relacionada con la susceptibilidad del material y $\alpha_j$ es el campo coherente inyectado en la guía $j$ (ver Figura 1). La cantidad $\vec{C} = (C_1, \ldots, C_{N-1})$ es el perfil de acoplamiento, mientras que la amplitud y la fase del perfil de inyección corresponden a $\vec{\alpha} = (|\alpha_1|, \ldots, |\alpha_N|)$ y $\vec{\phi} = (\arg \alpha_1, \ldots, \arg \alpha_N)$ (Barral et al., 2020b), respectivamente. Adicionalmente, en este contexto se entiende por desorden a la variación aleatoria de un conjunto de parámetros representativos del arreglo con respecto a una configuración homogénea, por ejemplo, la variación de los elementos $C_j$ del perfil de acoplamiento $\vec{C}$ con respecto a la fuerza de acoplamiento $C_0$.

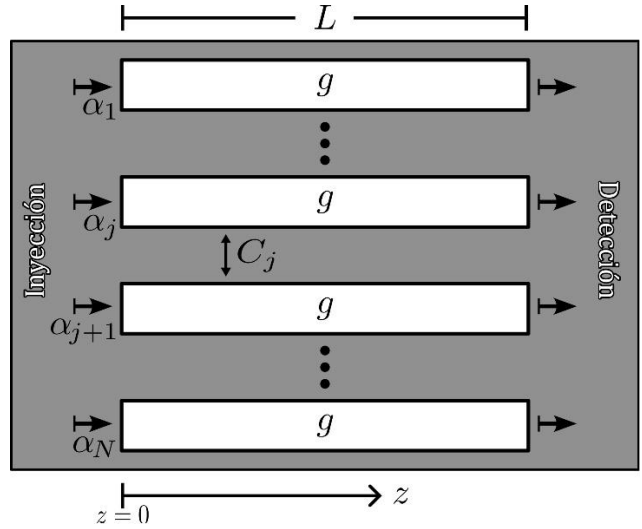

**Figura 1**. Arreglo no lineal con susceptibilidad $\chi^{(2)}$ y constante no lineal $g$ en donde se realiza una inyección inicial de un campo coherente intenso $\alpha_j$ en la guía $j$. El perfil de acoplamiento entre guías de onda es $\vec{C} = (C_1, \ldots, C_{N-1})$. Los efectos no lineales ocurren en la región $0 < z < L$, llamada zona de interacción.

Un sistema fotónico en un estado puro se puede representar mediante una superposición lineal de estados de Fock $|\psi\rangle = \sum_{n=0}^{\infty} c_n |n\rangle$, donde $c_n$ es un número complejo y $|n\rangle$ corresponde al autoestado $|\psi_n\rangle$ con energía $E_n$ (Furusawa, 2015).

Además, se definen los operadores de creación $\hat{a}^\dagger$ y aniquilación $\hat{a}$, respectivamente, mediante su acción en los estados número, $\hat{a}^\dagger |n\rangle = \sqrt{n+1} |n+1\rangle$ y $\hat{a}|n\rangle = \sqrt{n}|n-1\rangle$, donde los coeficientes $\sqrt{n+1}$ y $\sqrt{n}$ son simplemente constantes de normalización. Se puede interpretar que el operador $\hat{a}^\dagger$ ($\hat{a}$) crea (destruye) un cuanto de energía en el sistema. También es importante definir el operador número $\hat{n} = \hat{a}^\dagger \hat{a}$ (Fox, 2006).

Estos operadores serán útiles para definir el valor esperado de un observable $\hat{O}$ (en este caso, algún observable electrodinámico), dado por

$$\langle \hat{O} \rangle = \langle \psi | \hat{O} | \psi \rangle, \tag{1}$$

donde $\langle \hat{O} \rangle$ es el promedio de los resultados al medir la propiedad asociada a $\hat{O}$ en un ensamble de sistemas con estado $|\psi\rangle$ (Fox, 2006).

La evolución espacial está dada por

$$-i\hbar \frac{d}{dz} |\psi(z)\rangle = \widehat{M} |\psi(z)\rangle, \tag{2}$$

con $\widehat{M}$ el operador de momento lineal en la representación de interacción dado por (Barral et al., 2020a):

$$\widehat{M} = \hbar \sum_{j=1}^{N} \left[ C_j \hat{A}_{j+1} \hat{A}_j^\dagger + \eta_j \hat{A}_j^{\dagger 2} \right] + C.H., \tag{3}$$

con $\hat{A}_j^\dagger$ ($\hat{A}_j$) es el operador de creación (aniquilación) correspondiente a la guía de onda $j$ y $C.H.$ significa conjugado hermítico. Físicamente, el primer término representa un evento en donde se destruye un fotón en la guía $j+1$ y se crea un fotón en la guía $j$ con una probabilidad proporcional a $C_j$ y el segundo está relacionado con la creación de pares mediante el proceso no lineal de conversión paramétrica descendente espontánea.

Como el interés principal es modificar los parámetros del arreglo para ciertas aplicaciones, es importante aclarar que algunos parámetros, como la longitud de la muestra $L$ y el número de guías de onda $N$, no pueden ser alterados posterior a la construcción del dispositivo, mientras que esto sí es posible para el perfil de inyección (Barral et al., 2020b). Aunque normalmente el perfil de acoplamiento $\vec{C}$ tampoco puede ser modificado, se ha logrado crear arreglos dinámicos en los que se puede controlar la interacción entre los elementos de forma electroóptica (Yang et al., 2024).

Para estudiar la propagación de la luz a lo largo del arreglo, se utiliza el valor esperado normalizado del número de fotones en la guía $j$, dado por $n_j = \langle \widehat{N}_j \rangle / \sum_{i=1}^{N} \langle \widehat{N}_i \rangle$, donde $\langle \widehat{N}_j \rangle = \langle \hat{A}_j^\dagger \hat{A}_j \rangle$ es el valor esperado del operador número de la guía $j$. La cantidad $n_j$ funciona como una distribución de probabilidad y permite calcular la desviación estándar $\sigma$ definida por (Fox, 2006):

$$\sigma = \sqrt{\sum_{j=1}^{N} n_j M_j^2 - \left( \sum_{j=1}^{N} n_j M_j \right)^2}, \tag{4}$$

donde $M_j = 1, 2, \ldots, N-1, N$ es la posición relativa de la guía $j$. Cuando se inyecta solamente una guía de onda, esta desviación estándar se puede interpretar como la dispersión de la probabilidad de encontrar un fotón conforme se propaga la luz. Dicha cantidad sería igual a cero si solo es posible encontrar el fotón en la guía de onda inyectada inicialmente, y su valor incrementa conforme la probabilidad de encontrarlo en otras guías de onda más lejanas aumenta.

El *participation ratio* $PR$ es una medida del grado de deslocalización de la función de onda y se define a continuación

$$PR = \left( \sum_{j=1}^{N} n_j^2 \right)^{-1}. \tag{5}$$

Cuando $PR = 1$, la función de onda se encuentra completamente localizada en una guía de onda, mientras que $PR = N$ indica deslocalización completa en el arreglo (Karamlou et al., 2022). Finalmente, la matriz de correlación $\Gamma_{q,r}$ dada por

$$\Gamma_{q,r} = \begin{cases} \hat{A}_q^\dagger \hat{A}_r^\dagger \hat{A}_r \hat{A}_q, & q \neq r, \\ \hat{A}_q^{\dagger 2} \hat{A}_q^{\dagger 2}, & q = r, \end{cases} \tag{6}$$

corresponde a la probabilidad de, cuando se tiene un par de fotones indistinguibles, detectar un fotón en la guía $q$ y detectar el otro en la guía $r$.

## II. MATERIALES Y MÉTODOS

En arreglos con pocas guías de onda, la propagación de la luz se puede resolver numéricamente en Python mediante el uso de la librería QuTiP (Johansson et al., 2013), en donde se utiliza el punto de vista de Schrödinger, según la ecuación (2). No obstante, este método se vuelve computacionalmente demandante conforme crece el tamaño del sistema. Por este motivo, se limitó tanto el tamaño del arreglo como la dimensión del espacio de Hilbert asociado a cada modo a un rango donde el tiempo de simulación sea menor a unas cuantas horas.

Además, con el fin de estudiar solamente un par de fotones, se asumió que las condiciones son tales que ocurre un único evento no lineal, lo cual se consigue con una inyección de baja intensidad. Una vez obtenidas las soluciones, se caracterizó al sistema mediante los parámetros definidos en la Sección II.

### Implementación del desorden

El desorden se introdujo de forma similar al método utilizado por Karamlou et al. (2022), quienes modifican la energía de forma aleatoria en una red de enlace fuerte. En este caso, se consideraron tres tipos de desorden producidos. Es decir, en el perfil de acoplamiento $\vec{C}$, el perfil de amplitud de inyección $\vec{\alpha}$ y el perfil de fase de inyección $\vec{\phi}$. Los elementos de cada perfil se generaron aleatoriamente a partir de las siguientes distribuciones uniformes:

$$C_j \in [C_0(1-\kappa), C_0(1+\kappa)[ \tag{7}$$

$$\alpha_j \in [\alpha_0(1-\kappa), \alpha_0(1+\kappa)[ \tag{8}$$

$$\phi_j \in [0, 2\pi\kappa[ \tag{9}$$

donde el parámetro $\kappa$ es una medida de la intensidad del desorden; este es nulo cuando $\kappa = 0$ y $\kappa = 1$ indica el límite de desorden permitido físicamente por el sistema, ya que para valores mayores los elementos del perfil de acople y amplitud pueden tomar valores negativos. En el caso del perfil de fase, $\kappa > 1$ generaría una diferencia de fase redundante, es decir, mayor a $2\pi$.

**Configuración de las simulaciones**

Para todas las simulaciones, la dimensión del espacio de Hilbert de cada modo individual fue $m = 3$ y se trabajó con un arreglo de 9 guías de onda. Además, se utilizó $C_0 = 250$ m$^{-1}$, $g = 70$ W$^{-1/2}$/m, $\alpha_0 = \sqrt{5 \times 10^{-4}}$ W$^{1/2}$ y $L = 80$ cm, donde los primeros dos parámetros corresponden a valores experimentales típicos (Barral et al., 2021). Para cada intensidad de desorden dada, el valor de todas las cantidades físicas se calculó como el promedio de al menos 150 simulaciones o del número necesario para obtener un error porcentual menor al 0,75% entre el último y penúltimo promedio. Además, el promedio espacial de una cantidad $x$, denotado mediante $\overline{x}$, se calculó en el intervalo $10 < C_0 z < 20$, región en donde las cantidades alcanzaron un estado cercano a estacionario. En primer lugar, se introdujo desorden en el perfil de acoplamiento cuando solamente se inyecta una guía de onda para estudiar el comportamiento del *random walk*. Luego, se analizó la propagación cuando se inyectan todas las guías de onda, pero introduciendo desorden en los otros perfiles (amplitud y fase de inyección). Finalmente, se compararon algunas correlaciones cuánticas bajo la presencia de desorden en el perfil de acoplamiento.

## III. RESULTADOS Y DISCUSIÓN

### *Quantum random walk*

En primer lugar, se exploró cómo se ve afectada la propagación de la luz cuando se inyecta solo una guía de onda en un arreglo de 9 guías de onda y se introduce desorden en el perfil de acoplamiento $\vec{C}$. En la Figura 2 se muestra el comportamiento de un *quantum random walk* durante la propagación descrita por el parámetro $C_0 z$ con $z \in [0, L]$ cuando se inyecta la guía de onda central y cuando se inyecta la esquina (primera guía de onda) del arreglo. Cualitativamente, aumentar el desorden tiende a amortiguar la propagación lateral de la probabilidad, es decir, concentra la probabilidad de medir un fotón en la guía de onda inyectada, la cual se extingue rápidamente hacia los lados. Además, al aumentar el desorden se observa una disminución de las reflexiones en las paredes del arreglo.

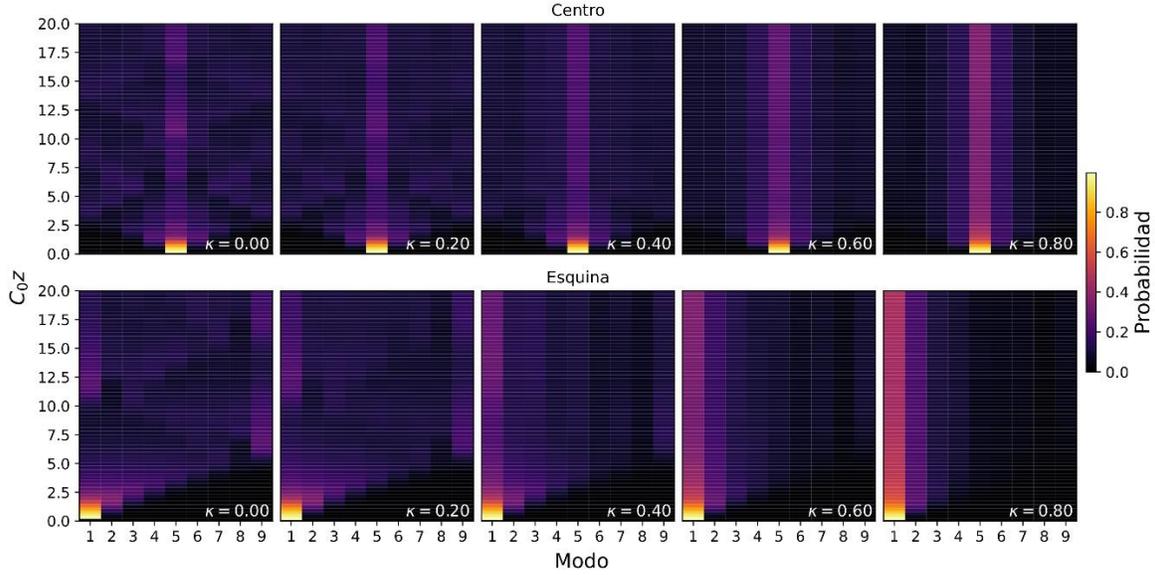

**Figura 2**. Comportamiento de un *quantum random walk* durante la propagación descrita por $C_0 z$ al introducir desorden en el perfil de acoplamiento. La probabilidad de encontrar un fotón en cada guía de onda está descrita por la barra de colores a la derecha. Se muestran los casos cuando se inyecta la guía de onda del centro (paneles superiores) y cuando se inyecta la primera guía del arreglo (paneles inferiores) para intensidades de desorden desde $\kappa = 0$ hasta $\kappa = 0{,}8$.

Adicionalmente, en la Figura 3a se presenta el promedio espacial de la desviación estándar $\sigma$ y el *participation ratio* $PR$ definidos en la ecuación (4) y en la ecuación (5), respectivamente. En los dos casos, ambas cantidades decrecen, esto indica que la función de onda se vuelve tanto más localizada como menos dispersa conforme incrementa el desorden. Es interesante rescatar que se alcanza una menor desviación estándar cuando la guía de onda inyectada es la central, pero se logra un menor $PR$ cuando se inyecta la esquina del arreglo. Esto se puede entender de la siguiente forma: cuando se inyecta la esquina del arreglo, la luz se puede "desplazar" 8 guías de onda hacia la derecha, mientras que solo se puede "desplazar" 4 guías de onda (a la izquierda o a la derecha) cuando se inyecta la guía de onda central; esto permite que se alcancen desviaciones estándar mayores en el primer caso, incluso cuando el desorden es nulo. El comportamiento del $PR$ posee una explicación similar. Conforme aumenta el desorden, la probabilidad de medir un fotón se extingue rápidamente al alejarse de la guía de onda inyectada inicialmente: cuando corresponde a la esquina, esta "difusión" de probabilidad solo ocurre hacia la derecha, pero sucede en ambas direcciones cuando se trata de la guía de onda central. Esto da origen a que la función de onda esté más localizada en el primer caso.

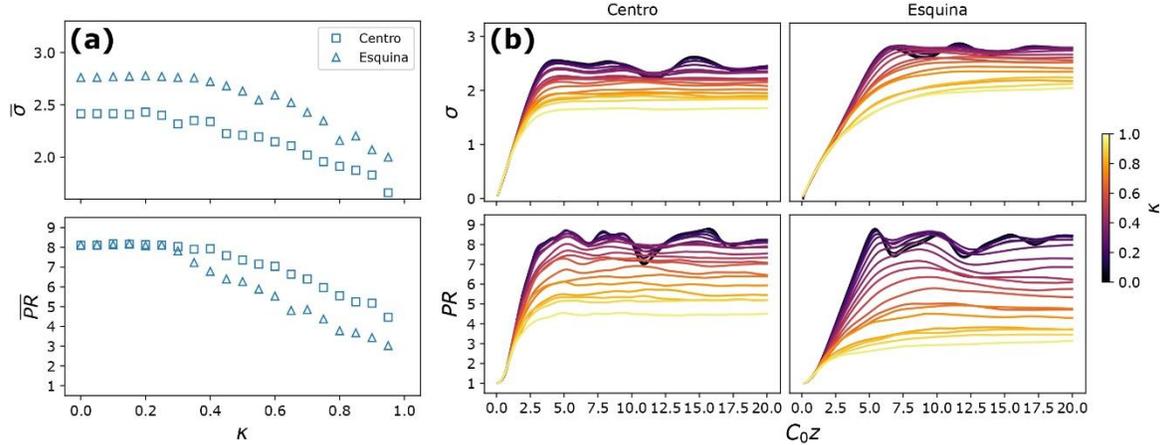

**Figura 3**. (a) Promedio espacial de la desviación estándar $\sigma$ y el *participation ratio* $PR$ para una inyección en la guía de onda central (cuadrados) y en la primera guía de onda (esquina) del arreglo (triángulos). (b) Desviación estándar $\sigma$ (páneles superiores) y *participation ratio* $PR$ (páneles inferiores) durante la propagación descrita por $C_0 z$. En (b), la columna de la izquierda corresponde a una inyección en la guía de onda central, mientras que la columna derecha compete a una inyección en la primera guía de onda (esquina) del arreglo. El código de colores de la derecha representa el parámetro $\kappa$, el cual indica el nivel de desorden asociado a cada curva.

La variación de la desviación estándar $\sigma$ y el *participation ratio* $PR$ en función de $C_0 z$ se muestra en la Figura 3b. Para valores pequeños de $C_0 z$ (antes de que ocurran las reflexiones con las paredes del arreglo), la desviación estándar presenta una dependencia aparentemente lineal en $z$. No obstante, no se puede afirmar que se trate de una propagación balística, ya que esto requiere un estudio más detallado de la velocidad de propagación. Por otro lado, note que en ausencia de desorden realmente no se alcanza un estado estacionario en ningún caso, sino que las cantidades presentan un comportamiento oscilatorio. No obstante, conforme se incremente al desorden, las variaciones disminuyen y en algunas ocasiones se llega a un estado estacionario. Por ejemplo, cuando se inyecta la guía de onda central la desviación estándar parece alcanzar rápidamente un estado estacionario. Lo mismo ocurre para el $PR$, pero con variaciones mayores. Cuando se inyecta la esquina del arreglo, la desviación estándar alcanza un estado estacionario hasta el final del rango estudiado de $C_0 z$, mientras que el $PR$ solo se estabiliza para las mayores intensidades de desorden. Analizar la variación en función de la propagación de cualquiera de estas cantidades es importante si interesa una convergencia rápida.

**Tipos de desorden**

Posteriormente, se estudiaron otros dos tipos de desorden, el generado en el perfil de amplitud $\vec{\alpha}$ y en el perfil de fase $\vec{\phi}$. En la Figura 4 se compara el promedio espacial del *participation ratio* en función del parámetro $\kappa$ para los diferentes tipos de desorden estudiados en este trabajo—es decir, en el acople, en la amplitud y en la fase—considerando un perfil de amplitud de inyección homogéneo. Cuando se inyecta más de una guía de onda, la desviación estándar no es necesariamente un parámetro representativo de la distribución de probabilidad en el arreglo y por lo tanto solo se analiza el *participation ratio*. Se puede observar que el $\overline{PR}$ es prácticamente independiente de la intensidad de desorden en el perfil de acoplamiento. Esto sugiere que imperfecciones en la construcción del arreglo de guías de onda no afectarían la deslocalización de la función de onda. Contrariamente, el desorden en el perfil de amplitud y el perfil de fase incrementan el $\overline{PR}$, teniendo un mayor impacto en el segundo caso. De hecho, cuando se introduce desorden en el perfil de fase el $\overline{PR}$ incrementa con notable mayor rapidez, alcanza valores superiores a 8 incluso con $\kappa = 0.5$ y tiende a 9 conforme aumenta la intensidad del desorden, en cuyo caso la función de onda está completamente deslocalizada.

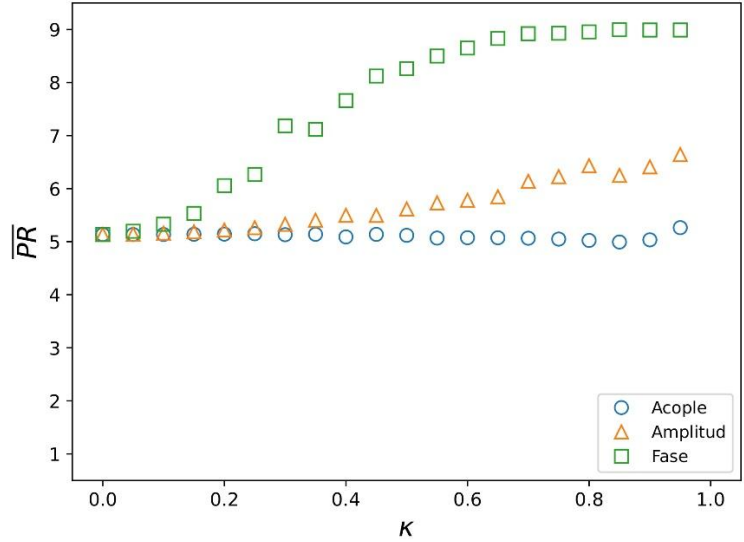

**Figura 4**. Promedio espacial del *participation ratio* en función de $\kappa$ para desorden en el perfil de acople $\vec{C}$ (círculos azules), el perfil de amplitud $\vec{\alpha}$ (triángulos anaranjados) y el perfil de fase $\vec{\phi}$ (cuadrados verdes).

### Correlaciones cuánticas

Finalmente, se estudió cuál es el efecto del desorden en la matriz de correlación definida en la ecuación (6), calculada al final de la propagación estudiada en este trabajo, es decir, en $C_0 z = 20$. La forma de la matriz de correlación en general varía para cada configuración, pero en ocasiones se producen patrones interesantes en donde varios de sus elementos son nulos.

Un caso relevante es cuando se tiene una amplitud de inyección homogénea y se varía la diferencia entre la fase de las guías de onda impares $\phi_i$ y la fase de las guías de onda pares $\phi_p$. El efecto del desorden en el perfil de acoplamiento se muestra en la Figura 5. Cuando $\phi_i - \phi_p = 0$ y no se ha implementado desorden, todos los elementos de $\Gamma_{q,r}$ donde $q$ o $r$ es par son nulos. Esta característica se mantiene conforme el desorden aumenta, pero también se observa que las probabilidades relativas de los elementos no nulos cambian, tal que para una intensidad de desorden $\kappa = 0,75$ es más probable encontrar ambos fotones en las esquinas del arreglo (guías de onda 1 y 9). Note que esto difiere de cuando no hay desorden, en cuyo caso es ligeramente menos probable encontrar a ambos fotones en la misma guía de onda, es decir, los fotones tienden a aglutinarse con el desorden. Por otro lado, cuando $\phi_i - \phi_p = \pi$, todos los elementos de la matriz de correlación se mantienen invariantes al incrementar la intensidad de desorden.

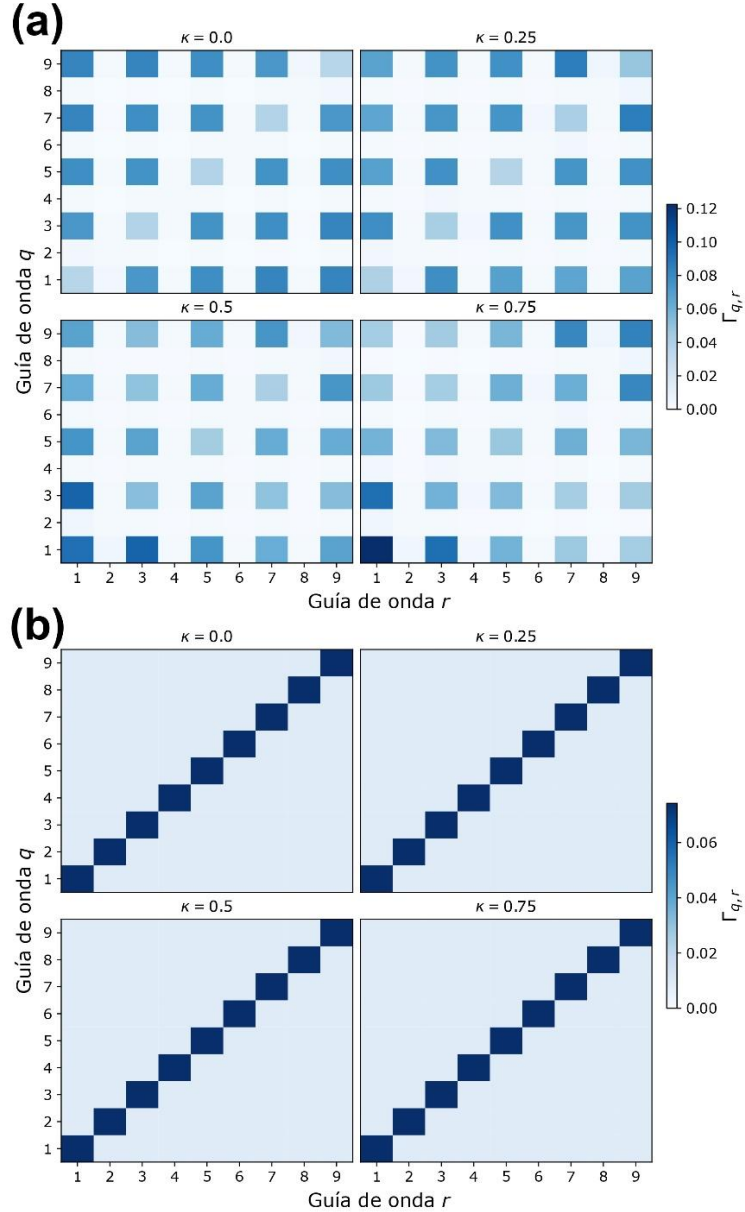

**Figura 5**. Matriz de correlación para distintas intensidades de desorden $\kappa$ cuando se introduce desorden en el perfil de acoplamiento. La barra azul denota la magnitud de cada elemento en unidades arbitrarias, en donde el tono claro de la izquierda de la barra corresponde a entradas nulas. En este caso, la magnitud de las entradas de la matriz de correlación aumenta conforme los tonos se tornan más oscuros. Se muestra los casos en donde la diferencia entre la fase de las guías de onda impares $\phi_i$ y la fase de las guías de onda pares $\phi_p$ es (a) $\phi_i - \phi_p = 0$ y (b) $\phi_i - \phi_p = \pi$.

Esto demuestra que, para los dos casos particulares estudiados, las correlaciones cuánticas entre los fotones son robustas ante el desorden, en especial aquellas que son nulas. Sin embargo, sí se observan diferencias entre los valores relativos de las probabilidades no nulas, lo cual puede alterar dónde es más probable detectar los fotones.

## IV. CONCLUSIONES

En el estudio del *quantum random walk*, se determinó que la presencia de desorden disminuye la dispersión y aumenta la localización de la función de onda, siendo el primer efecto mayor cuando se inyecta la guía de onda central y el segundo para una guía de onda en la esquina del arreglo. Además, se observó que la propagación tiende a ser balística, lo cual es característico de arreglos de guías de onda. Contrariamente, la implementación de desorden en el perfil de amplitud de inyección y fase de inyección provocó un incremento en la deslocalización de la función de onda.

Por otro lado, para los casos específicos estudiados en este trabajo, al analizar los cambios en la matriz de correlación conforme la intensidad de desorden se vuelve mayor, se observó que la distribución de probabilidad es robusta ante el desorden, en particular la correspondiente a los elementos nulos de la matriz. Los resultados obtenidos pueden ser de utilidad para el desarrollo de las tecnologías cuánticas que se basen en arreglos de guías de onda no lineales.

## V. REFERENCIAS

Anderson, P. W. (1958). Absence of diffusion in certain random lattices. *Physical Review*, *109*(5), 1492–1505.

Bai, Y. F., Xu, P., Lu, L. L., Zhong, M. L. y Zhu, S. N. (2016). Two-photon Anderson localization in a disordered quadratic waveguide array. *Journal of Optics*, *18*(5), 055201.

Barral, D., D'Auria, V., Doutre, F., Lunghi, T., Sébastien Tanzilli, Rambu, A. P., Sorin Tascu, Levenson, J. A., Belabas, N. y Kamel Bencheikh. (2021). Supermode-based second harmonic generation in a nonlinear interferometer. *Optics Express*, *29*(23), 37175–37175.

Barral, D., Mattia Walschaers, Kamel Bencheikh, Parigi, V., Levenson, J. A., Treps, N. y Belabas, N. (2020a). Quantum state engineering in arrays of nonlinear waveguides. *Physical Review A*, *102*(4), 043706.

Barral, D., Mattia Walschaers, Kamel Bencheikh, Parigi, V., Levenson, J. A., Treps, N. y Belabas, N. (2020b). Versatile photonic entanglement synthesizer in the spatial domain. *Physical Review Applied*, *14*(4), 044025.

Blanco-Redondo, A., Bell, B., Oren, D., Eggleton, B. J. y Mordechai Segev. (2018). Topological protection of biphoton states. *Science*, *362*(6414), 568–571.

Bromberg, Y., Yoav Lahini, Morandotti, R. y Silberberg, Y. (2009). Quantum and classical correlations in waveguide lattices. *Physical Review Letters*, *102*(25), 253904.

Fox, M. (2006). *Quantum optics: an introduction*. Oxford University Press.

Furusawa, A. (2015). *Quantum states of light* (Vol. 10, pp. 69–100). Springer Japan.

Hamilton, C. S., Kruse, R., Sansoni, L., Silberhorn, C. y Jex, I. (2014). Driven quantum walks. *Physical Review Letters*, *113*(8), 083602.

Johansson, J. R., Nation, P. D. y Nori, F. (2013). QuTiP 2: A Python framework for the dynamics of open quantum systems. *Computer Physics Communications*, *184*(4), 1234–1240.

Karamlou, A. H., Jochen Braumüller, Yariv Yanay, Paolo, A. D., Harrington, P. M., Kannan, B., Kim, D., Morten Kjaergaard, Melville, A., Muschinske, S., Niedzielski, B. M., Antti Vepsäläinen, Winik, R., Yoder, J. L., Schwartz, M., Tahan, C., Orlando, T. P., Gustavsson, S. y Oliver, W. D. (2022). Quantum transport and localization in 1d and 2d tight-binding lattices. *Npj Quantum Information*, *8*(1), 35.

Kempe, J. (2003). Quantum random walks: An introductory overview. *Contemporary Physics*, *44*(4), 307–327.

Kokkinakis, E. T., Makris, K. G. y Economou, E. N. (2024). Anderson localization versus hopping asymmetry in a disordered lattice. *Physical Review A*, *110*(5), 053517.

Laurent Labonté, Olivier Alibart, d'Auria, V., Doutre, F., Etesse, J., Sauder, G., Martin, A., Éric Picholle y Sébastien Tanzilli. (2024). Integrated photonics for quantum communications and metrology. *PRX Quantum*, *5*(1).

Luo, W., Cao, L., Shi, Y., Wan, L., Zhang, H., Li, S., Chen, G., Li, Y., Li, S., Wang, Y., Sun, S.-H., Karim, M. F., Cai, H., Kwek, L.-C. y Liu, A. Q. (2023). Recent progress in quantum photonic chips for quantum communication and internet. *Light-Science & Applications*, *12*(175).

Perets, H. B., Yoav Lahini, Pozzi, F., Sorel, M., Morandotti, R. y Silberberg, Y. (2008). Realization of quantum walks with negligible decoherence in waveguide lattices. *Physical Review Letters*, *100*(17), 170506.

Peruzzo, A., Mirko Lobino, Jonathan, Matsuda, N., Politi, A., Poulios, K., Zhou, X.-Q., Yoav Lahini, Ismail, N., Wörhoff, K., Bromberg, Y., Silberberg, Y., Thompson, M. G. y OBrien, J. L. (2010). Quantum walks of correlated photons. *Science*, *329*(5998), 1500–1503.

Rai, A. y Angelakis, D. G. (2012). Dynamics of nonclassical light in integrated nonlinear waveguide arrays and generation of robust continuous-variable entanglement. *Physical Review A*, *85*(5), 052330.

Solntsev, A. S., Sukhorukov, A. A., Neshev, D. N. y Kivshar, Y. S. (2012). Spontaneous parametric down-conversion and quantum walks in arrays of quadratic nonlinear waveguides. *Physical Review Letters*, *108*(2), 023601.

Wang, J., Sciarrino, F., Laing, A. y Thompson, M. G. (2020). Integrated photonic quantum technologies. *Nature Photonics*, *14*(5), 273–284.

Yang, Y., Chapman, R. J., Haylock, B., Lenzini, F., Joglekar, Y. N., Mirko Lobino y Peruzzo, A. (2024). Programmable high-dimensional Hamiltonian in a photonic waveguide array. *Nature Communications*, *15*(50).

Zhang, C., Huang, Y., Liu, B., Li, C. y Guo, G. (2021). Spontaneous parametric Down-Conversion sources for multiphoton experiments. *Advanced Quantum Technologies*, *4*(5), 2000132.